\documentclass[twocolumn]{elsart5p}

\usepackage{mcite}
\usepackage{graphicx}
\usepackage{amsmath}
\usepackage{amssymb}
\usepackage{subfigure}

\newcommand{\bem}{\begin{pmatrix}}
\newcommand{\eem}{\end{pmatrix}}
\newcommand{\be}{\begin{equation}}
\newcommand{\ee}{\end{equation}}
\newcommand{\bbe}{\begin{equation*}}
\newcommand{\eee}{\end{equation*}}
\newcommand{\mbf}[1]{\mathbf{#1}}
\newcommand{\mr}[1]{\mathrm{#1}}
\newcommand{\bea}{\begin{eqnarray}}
\newcommand{\eea}{\end{eqnarray}}
\newcommand{\bbea}{\begin{eqnarray*}}
\newcommand{\eeea}{\end{eqnarray*}}
\newcommand{\dEdX}{w}
\newcommand{\Xmax}{\ensuremath X_\mathrm{max}}
\newcommand{\gcm}{g\,cm$^{-2}$}
\newcommand{\Ne}{\ensuremath N^\mathrm{e}}
\def\EeV{\ifmmode {\mathrm{\ Ee\kern -0.1em V}}\else
                   \textrm{Ee\kern -0.1em V}\fi}%
\def\eV{\ifmmode {\mathrm{\ e\kern -0.1em V}}\else
                   \textrm{e\kern -0.1em V}\fi}%

\begin{document}

\begin{frontmatter}

\title{Reconstruction of Longitudinal Profiles of Ultra-High Energy 
       Cosmic Ray Showers from Fluorescence and Cherenkov Light Measurements}

\author[fzk]{M.\ Unger\corauthref{cor1}},
\author[uoa]{B.R.\ Dawson},
\author[fzk]{R.\ Engel},
\author[fzk]{F.\ Sch\"ussler} and 
\author[fzk]{R.\ Ulrich}

\corauth[cor1]{corresponding author, Michael.Unger@ik.fzk.de}
\address[fzk]{Institut f\"ur Kernphysik, Forschungszentrum Karlsruhe, Postfach 3640, 76021~Karlsruhe, Germany}
\address[uoa]{Department of Physics, University of Adelaide, Adelaide 5005, Australia}

\begin{abstract}
  We present a new method for the reconstruction of the longitudinal
  profile of extensive air showers induced by ultra-high energy cosmic
  rays. In contrast to the typically considered shower size profile,
  this method employs directly the ionization energy deposit of the
  shower particles in the atmosphere.  Due to universality of the
  energy spectra of electrons and positrons, both fluorescence and
  Cherenkov light can be used simultaneously as signal to infer the
  shower profile from the detected light. The method is based on an
  analytic least-square solution for the estimation of the shower
  profile from the observed light signal. Furthermore, the
  extrapolation of the observed part of the profile with a
  Gaisser-Hillas function is discussed and the total statistical
  uncertainty of shower parameters like total energy and shower
  maximum is calculated.
\end{abstract}
\begin{keyword}
  Cosmic rays \sep extensive air showers \sep air shower reconstruction
  \sep air fluorescence \sep Cherenkov light 
\PACS 96.50.sd
\end{keyword}
\end{frontmatter}

\section{Introduction}

The particles of an extensive air shower excite nitrogen molecules in
the atmosphere, which subsequently radiate ultraviolet
fluorescence light isotropically. This fluorescence light signal can be measured
with appropriate optical detectors such as the fluorescence telescopes
of HiRes~\cite{fd:HiRes}, the Pierre Auger Observatory~\cite{fd:Auger}
or the Telescope Array~\cite{fd:TA}.

The number of emitted fluorescence photons is expected to be
proportional to the energy deposited by the shower particles. Recent
measurements of the fluorescence yield in the laboratory confirm this
expectation within the experimental
uncertainties~\cite{Kakimoto:1995pr,Waldenmaier:2007am,airflydEdX}. 
Non-radiative
processes of nitrogen molecule de-excitation lead to a temperature,
pressure and humidity dependence of the fluorescence yield (see e.g.~\cite{airflyAtm}). For
atmospheric parameters of relevance to the reconstruction of air
showers of ultra-high energy cosmic rays, the pressure dependence of
the ionization energy deposit per meter track length of a charged
particle is almost perfectly canceled by the pressure dependence of
the fluorescence yield (see, for example,
\cite{Keilhauer:2005nk}). 
Therefore only a weak pressure and
temperature dependence has to be taken into account if the number of
emitted photons is converted to a number of charged particle times
track length, as has been done in the pioneering Fly's Eye experiment
\cite{Baltrusaitis:1985mx}. The reconstructed longitudinal shower
profile is then given by the number of charged particles as function
of atmospheric depth.

The approximation of assuming a certain number of fluorescence photons
per meter of charged particle track and the corresponding expression
of the longitudinal shower development in terms of shower size are
characterized by a number of conceptual shortcomings. Firstly the
energy spectrum of particles in an air shower changes in the course of
its development. A different rate of fluorescence photons per charged
particle has to be assumed for early and late stages of shower
development as the ionization energy deposit depends on the particle
energy \cite{Song:1999wq}. Secondly the tracks of low-energy particles
are not parallel to the shower axis leading to another correction that
has to be applied \cite{Alvarez-Muniz:2003my}. Thirdly the quantity
``shower size'' is not suited to a precise comparison of measurements
with theoretical predictions. In air shower simulations, shower size
is defined as the number of charged particles above a given energy
threshold $E_{\rm cut}$ that cross a plane perpendicular to the shower
axis. Setting this threshold very low to calculate the shower size with an
accuracy of $\sim 1$\% leads to very large simulation times as the
number of photons diverges for $E_{\rm cut}\rightarrow
0$. Moreover, the shower size reconstructed from data depends on
simulations itself since the shower size is not directly related to the
fluorescence light signal.

These conceptual problems can be avoided by directly using energy
deposit as the primary quantity for shower profile reconstruction as well
as comparing experimental data with theoretical predictions. Due to
the proportionality of the number of fluorescence photons to the
energy deposit, shower simulations are not needed to reconstruct the
total energy deposit at a given depth in the atmosphere. Another
advantage is that the calorimetric energy of the shower is directly
given by the integral of the energy deposit profile \cite{linsleydEdX}. 
Furthermore the
energy deposit profile is a well-defined quantity that can be
calculated straight-forwardly in Monte Carlo simulations and does not depend on
the simulation threshold \cite{Risse:2003fw}.

Most of the charged shower particles travel faster than the speed of
light in air, leading to the emission of Cherenkov light. Thus, in
general, the optical signal of an air shower consists of both
fluorescence and Cherenkov light contributions. In the traditional
method \cite{Baltrusaitis:1985mx} for the reconstruction of the
longitudinal shower development, the Cherenkov light is iteratively
subtracted from the measured total light. The drawbacks of this method
are the lack of convergence for events with a large amount of
Cherenkov light and the difficulty of propagating the uncertainty of
the subtracted signal to the reconstructed shower profile.

An alternative procedure, used in \cite{Abbasi:2004nz}, 
is to assume a functional form for the longitudinal development of the shower, 
calculate the corresponding light emission and vary the parameters of the
shower curve until a satisfactory agreement with the observed light at the
detector is obtained. Whereas in this scheme the convergence problems 
of the aforementioned method are avoided, its major disadvantage is
that it can only be used if the showers indeed follow the functional form 
assumed in the minimization. 

It has been noted in~\cite{cher:giller} that, due to the universality
of the energy spectra of the secondary electrons and positrons within an
air shower, there exists a non-iterative solution for the reconstruction
of a longitudinal shower profile from light detected by fluorescence telescopes.

Here we will present an analytic least-square solution for the
estimation of the longitudinal energy deposit profile of air showers
from the observed light signal, in which both fluorescence and
Cherenkov light contributions are treated as signal. We will also
discuss the calculation of the statistical uncertainty of the shower
profile, including bin-to-bin correlations. Finally we will introduce
a constrained fit to the detected shower profile for extrapolating it
to the regions outside the field of view of the fluorescence
telescope. This constrained fit allows us to always use the full set
of profile function parameters independent of the quality of the
detected shower profile.

\begin{figure*}
 \begin{center}
 \includegraphics[clip,width=0.95\textwidth]{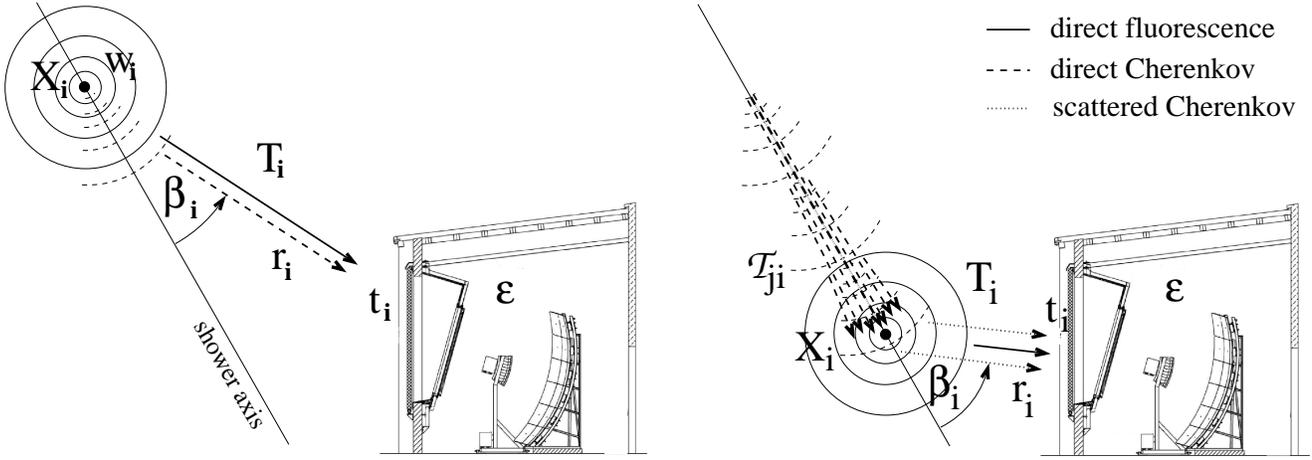}
 \caption{Illustration of the isotropic fluorescence light
          emission (solid circles), Cherenkov beam along the shower axis (dashed arcs) and the 
          direct (dashed lines) and scattered (dotted lines) Cherenkov light contributions.}\label{fig1}
 \end{center}
\end{figure*}

\section{Fluorescence and Cherenkov Light Signals}

The non-scattered, i.e.\ directly observed fluorescence light emitted at a certain slant depth $X_i$ is measured
at the detector at a time $t_i$. Given the fluorescence yield $Y^\mr{f}_i$ 
\cite{Kakimoto:1995pr,Nagano:2004am,airfly,Waldenmaier:2007am} 
at this point of the atmosphere, the number of photons produced at the shower in a slant
depth interval $\Delta X_i$ is 
\be
N_\gamma^\mr{f}(X_i) =  Y^\mr{f}_i\,\dEdX_i\,\Delta X_i.
\label{eq:fluo1}
\ee
Here, $\dEdX_i$ denotes the energy deposited per unit depth at slant depth $X_i$ (cf.\ Fig.~\ref{fig1}) 
and is defined as 
\be
\dEdX_i = \frac{1}{\Delta X_i} \int_0^{2 \pi} {\rm d}\varphi \int_0^\infty r {\rm d}r 
\int_{\Delta z_i} {\rm d}z \frac{{\rm d}E_{\rm dep}}{{\rm d}V},
\ee
where ${\rm d}E_{\rm dep}/{\rm d}V$ is the energy deposit per unit volume and ($\varphi,R,z$) 
are cylinder coordinates with the shower axis at $R=0$. The distance interval 
$\Delta z_i$ along the shower axis is given by the slant depth interval $\Delta X_i$. 
The fluorescence yield $Y^\mr{f}_i$ is the number of photons expected per unit 
deposited energy for the atmospheric pressure and temperature at slant depth $X_i$.
The photons from Eq.~(\ref{eq:fluo1}) are distributed over
a sphere with surface $4\,\pi\,r_i^2$, where $r_i$ denotes the distance of the detector. 
Due to atmospheric attenuation only a fraction $T_i$ of them reach the detector aperture with area $A$. Given
 a light detection efficiency of $\varepsilon$, the measured fluorescence light flux 
$y_i^\mr{f}$ can be written as
\be
  y_i^\mr{f}=  d_i\,Y^\mr{f}_i \,\dEdX_i\,\Delta X_i,
  \label{eq:fluo}
\ee
where the abbreviation
$
  d_i=\varepsilon\, T_i \frac{A}{4\,\pi\,r_i^2} 
$ 
is used. For the sake of clarity
the wavelength dependence of $Y$, $T$ and $\varepsilon$ 
will be disregarded in the following, but  discussed later.\\
The number of Cherenkov photons emitted at the shower is proportional to the 
number of charged particles above the Cherenkov threshold energy. Since the 
electromagnetic component dominates the shower development, the emitted Cherenkov light, 
$N_\gamma^\mr{C}$, can be calculated from 
\be
N_\gamma^\mr{C}(X_i) = Y^\mr{C}_i\,\Ne_i\,\Delta X_i,
\ee
where $\Ne_i$ denotes the number of electrons and positrons above a certain energy cutoff, which
is constant over the full shower track and not to be confused with the Cherenkov emission
energy threshold. Details of the Cherenkov light
production like these thresholds are included
in the Cherenkov yield factor $Y^\mr{C}_i$ \cite{cher:giller,cher:hillasangle,cher:hillaslongi,cher:nerling}. 

Although 
Cherenkov photons are emitted in a narrow cone along the particle direction, they 
cover a considerable angular range with respect to the shower axis, because
the charged particles are deflected from the primary particle direction due to multiple
scattering. Given the fraction $f_\mr{C}(\beta_i)$ of Cherenkov photons  per solid angle emitted at an angle 
$\beta_i$ with respect
to the shower axis~\cite{cher:hillasangle,cher:nerling}, the light flux at the detector 
aperture originating from direct Cherenkov light is
\be
  y_i^\mr{Cd}= d_i\, f_\mr{C}(\beta_i)\,Y^\mr{C}_i\,\Delta X_i \, \Ne_i.
  \label{eq:dcher}
\ee
Due to the forward peaked nature of Cherenkov light production, an intense Cherenkov
light beam builds up along the shower as it traverses the atmosphere (cf.~Fig.~\ref{fig1}).
If a fraction $f_\mr{s}(\beta_i)$ of the beam is scattered towards the observer it can
contribute significantly to the total light received at the detector. In a simple one-dimensional model
the number of photons in the beam at depth~$X_i$ is just the sum of Cherenkov light produced at 
all previous depths $X_j$ attenuated on the way from $X_j$ to $X_i$ by $\mathcal{T}_{ji}$:
\be
N_\gamma^\mr{beam}(X_i) = \sum_{j=0}^i \mathcal{T}_{ji}\,Y^\mr{C}_j\,\Delta X_j\,\Ne_j.
\ee
Similar to the direct contributions, the scattered Cherenkov light received at the detector is then
\be
  y_i^\mr{Cs}= d_i\, f_\mr{s}(\beta_i)\,\sum_{j=0}^i \mathcal{T}_{ji}\,Y^\mr{C}_j\,\Delta X_j\,\Ne_j.
  \label{eq:scher}
\ee
Finally, the total light received at the detector at the time $t_i$ is obtained by 
adding the scattered and direct
light contributions:
\be
   y_i=y_i^\mr{f}+y_i^\mr{Cd}+y_i^\mr{Cs}.
\ee
\begin{figure*}[t!]
       \hspace{-1cm}
       \subfigure[Light at aperture.]{\label{fig:apertureExample}
            \includegraphics[width=0.49\linewidth] {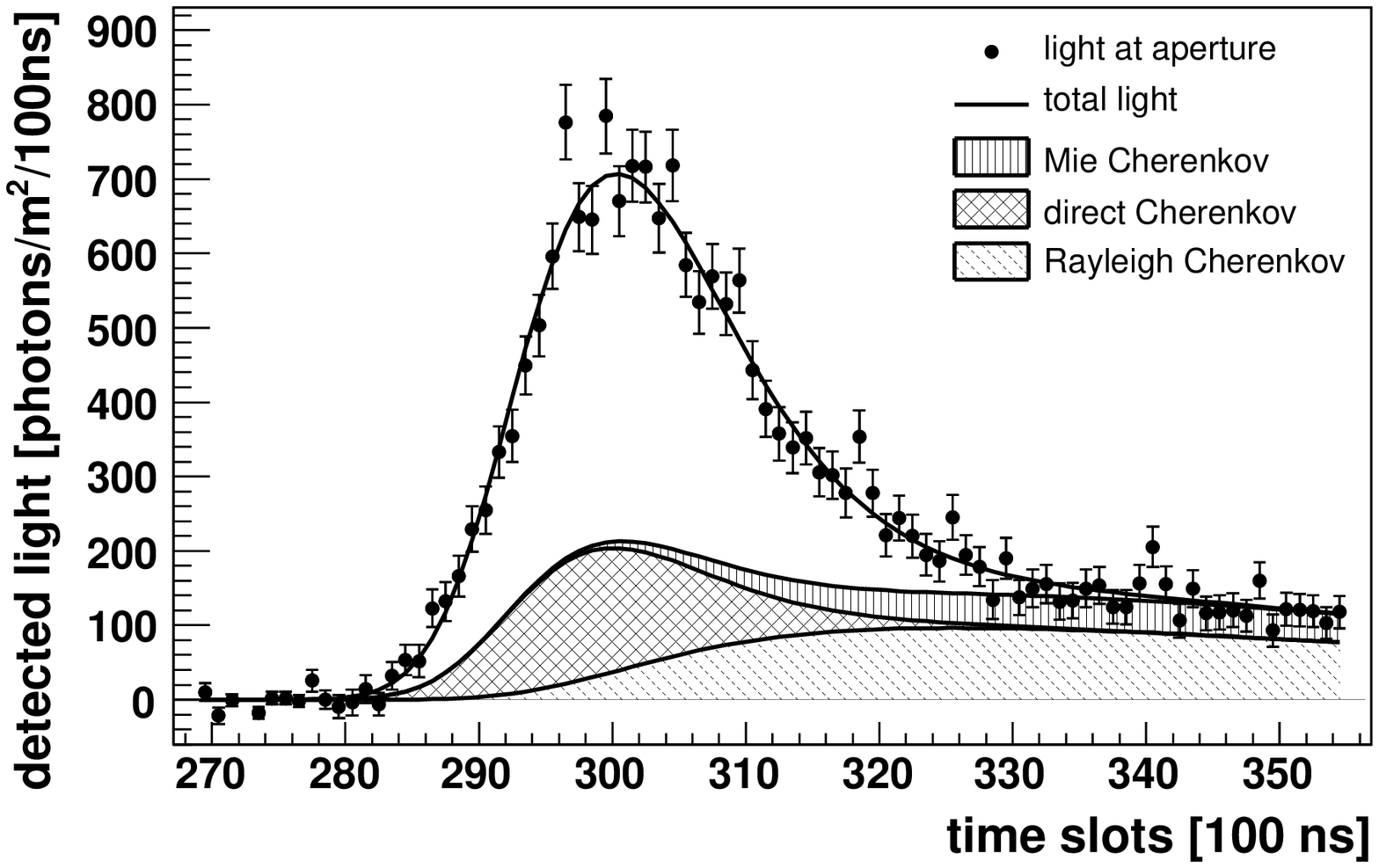}}
       \subfigure[Energy deposit profile]{\label{profileExample}
            \includegraphics[width=0.49\linewidth] {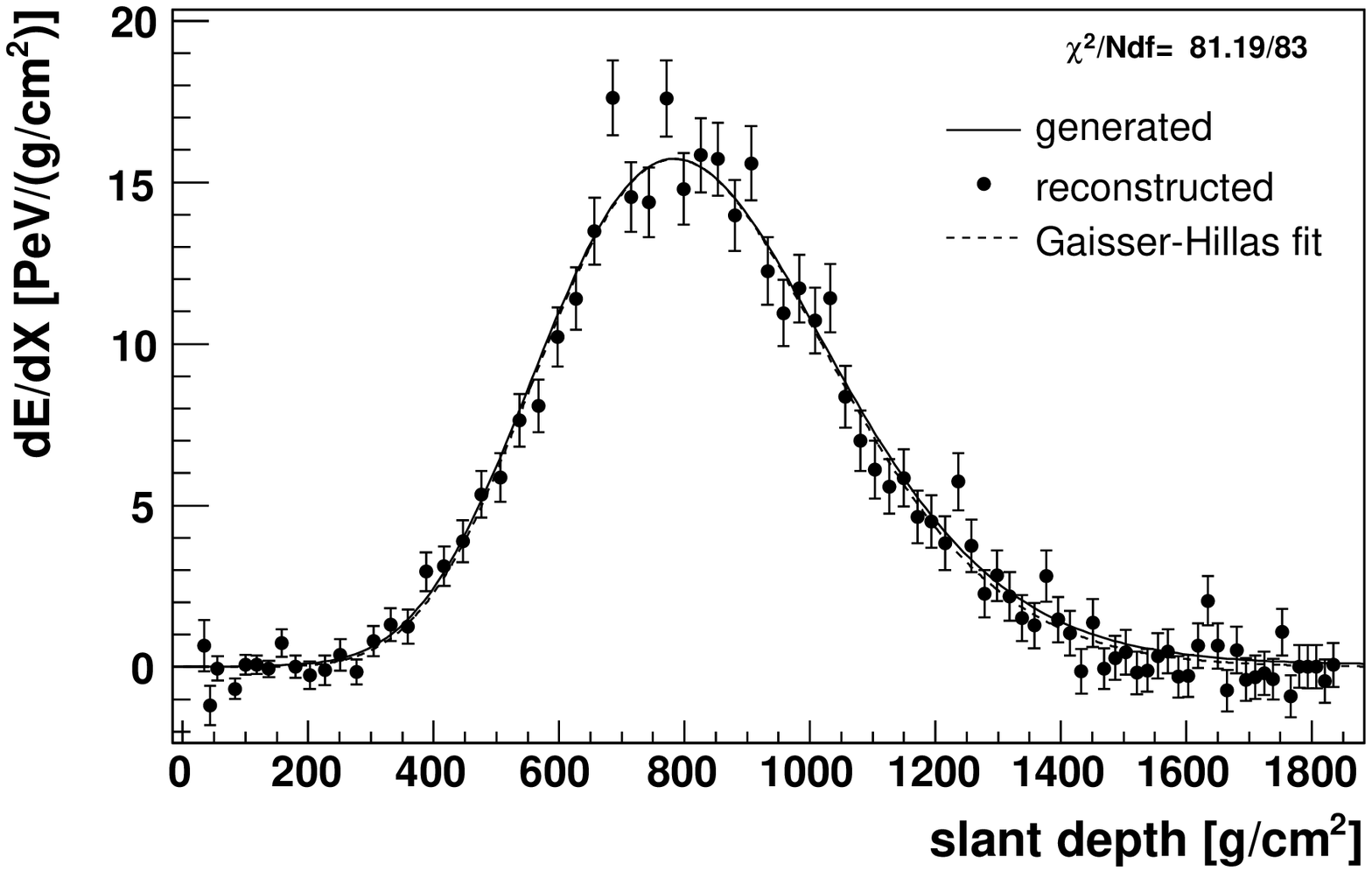}}
        \caption{Example of a simulated 10$^{19}$ \eV proton shower.}
        \label{fig:simulationExample}
\end{figure*}

\section{Analytic Shower Profile Reconstruction}
\label{sec:reco}
The aim of the profile reconstruction is to estimate the energy deposit and/or 
electron profile from the light flux observed at the detector. At
first glance this seems to be hopeless, since at each depth there are the
two unknown variables $\dEdX_i$ and $\Ne_i$, and only one measured quantity,
namely $y_i$.
Since the total energy deposit is just the sum of the energy loss of electrons,
$\dEdX_i$ and $\Ne_i$ are related via
\be
  \dEdX_i=\Ne_i \, \int_0^\infty f_\mr{e}(E,X_i) \;\dEdX_\mr{e}(E)\; \mr{d}E,
  \label{eq:nededx}
\ee
where $f_\mr{e}(E,X_i)$ denotes the normalized electron energy distribution 
and $\dEdX_\mr{e}(E)$ is the energy loss per unit depth of a single electron with energy $E$.
As is shown in~\cite{cher:giller,cher:hillaslongi,cher:nerling}, the electron energy 
spectrum $f_\mr{e}(E,X_i)$ is universal in shower age $s_i=3/(1+2\Xmax/X_i)$,
 i.e.~it does not depend on the primary mass or energy, but only on
the relative distance to the shower maximum, $\Xmax$. Eq.~(\ref{eq:nededx}) can
thus be simplified to
\be
   \dEdX_i=\Ne_i \; \alpha_i.
   \label{eq:dne}
\ee
where $\alpha_i$ is the average energy deposit per unit depth per electron at shower age 
$s_i$. Parameterizations of $\alpha_i$ can be found in~\cite{Song:1999wq,cher:nerling}.
 With this one-to-one 
relation (Eq.~\ref{eq:dne}) between the energy deposit and the number of electrons, the shower profile
is readily calculable from the equations given in the last section.
For the solution of the problem, it is convenient to rewrite the 
relation between energy deposit and
light at the detector
in matrix notation:
Let 
$
\mbf{y} = 
( y_1, y_2, \dots, y_n )^\mr{T}
$ 
 be the $n$-component vector (histogram) of the measured photon flux  at the aperture and
$
\mbf{\dEdX} = 
( \dEdX_1, \dEdX_2, \dots, \dEdX_n )^\mr{T}
$ 
the energy deposit vector at the shower track. Using the expression
\be
   \mbf{y}=\mbf{C}\mbf{\dEdX}
   \label{eq:matrix}
\ee
the elements of the {\itshape Cherenkov-fluorescence matrix} $\mbf{C}$ can be found
by a comparison with the coefficients in equations~(\ref{eq:fluo}), (\ref{eq:dcher}) 
and (\ref{eq:scher}):
\be
C_{ij}=
\begin{cases}
   0,\hfill i<j\!\;\\
   c_{i}^\mr{d}+c_{ii}^\mr{s},\;\;\; i=j\,\\
   c_{ij}^\mr{s},\hfill i>j,\\
\end{cases}
\label{eq:elements}
\ee
where
\be
c_{i}^\mr{d}=d_i \left(Y^\mr{f}_i + f_\mr{C}(\beta_i)\,Y^\mr{C}_i/\alpha_i\right)\,\Delta X_i
\ee
and
\be
c_{ij}^\mr{s}=d_i\,  f_\mr{s}(\beta_i)\,\mathcal{T}_{ji}\,Y^\mr{C}_j/\alpha_j\,\Delta X_j .
\ee
The solution of Eq.~(\ref{eq:matrix}) can be obtained by inversion, leading to the energy deposit
estimator $\widehat{\mbf{\dEdX}}$:
\be
    \widehat{\mbf{\dEdX}}=\mbf{C}^{-1}\mbf{y}\,.
\ee
Due to the triangular structure of the Cherenkov-fluorescence matrix the inverse can
be calculated quickly even for matrices with large dimension. As the matrix elements in Eq.~(\ref{eq:elements})
are always $\ge 0$, $\mbf{C}$ is never singular.\\
The statistical uncertainties of $\widehat{\mbf{\dEdX}}$ are obtained by error propagation:
\be
\mbf{V_\dEdX}=\mbf{C}^{-1} \,\mbf{V_y} \left(\mbf{C}^{\mr{T}}\right)^{-1}\;.
\ee  
It is interesting to note that even if the measurements $y_i$ are uncorrelated, i.e.~their 
covariance matrix $\mbf{V_y}$ is diagonal, the calculated energy loss values 
$\widehat{\dEdX}_i$ are not. This is 
because the light observed during time interval
$i$ does not solely originate from $\dEdX_i$, but  also receives 
a contribution from earlier shower parts $\dEdX_j$, $j<i$, 
via the 'Cherenkov light beam'.\\
\section{Wavelength Dependence}
\label{sec:lambda}
Until now it has been assumed that the shower induces light emission at a single wavelength $\lambda$. 
In reality, the fluorescence yield
shows distinct emission peaks and the number of Cherenkov photons produced is proportional
to $\frac{1}{\lambda^2}$. In addition the wavelength dependence of the
detector efficiency and the light transmission need to be taken into account.
Assuming that a binned wavelength distribution of the yields is available
($
   Y_{ik}=\int_{\lambda_k-\Delta \lambda}^{\lambda_k+\Delta \lambda} Y_{i}(\lambda)\, \mr{d}\lambda
$), 
the above considerations still hold when replacing $c_i^\mr{d}$ and
$c_{ij}^\mr{s}$ in Eq.~(\ref{eq:elements}) by
\be
\tilde{c}_i^\mr{\,d}=\Delta X_i\sum_k\,d_{ik} \left(Y^\mr{f}_{ik} +
                     f_\mr{C}(\beta_i)\,Y^\mr{C}_{ik}/\alpha_i\right)
\ee
and
\be
\tilde{c}_{ij}^\mr{\,s}=\Delta X_j\sum_k\,d_{ik}\, f_\mr{s}(\beta_i)\,
                        \mathcal{T}_{jik}\,Y^\mr{C}_{jk}/\alpha_j,
\ee
where
\be
  d_{ik}=\frac{\varepsilon_k\, T_{ik} }{4\,\pi\,r_i^2}.
\ee
The detector efficiency $\varepsilon_k$ and transmission coefficients $T_{ik}$ 
and $\mathcal{T}_{jik}$ are evaluated at the wavelength~$\lambda_k$.
\section{Validation with Air Shower Simulations}
\label{sec:performance}
\begin{figure}[t]
  \begin{center}
    \includegraphics[width=0.9\linewidth] {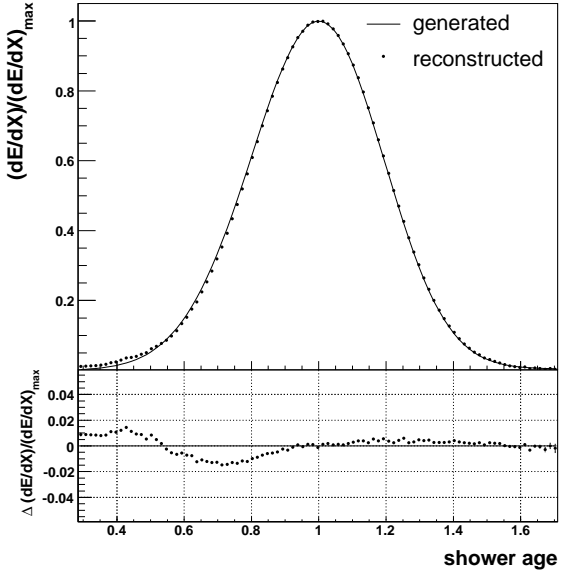}
        \caption{Average generated and reconstructed energy deposit profiles.}
        \label{fig:average}
   \end{center}
\end{figure}

In order to test the performance of the reconstruction algorithm we will use  in the following
simulated fluorescence detector data. 
For this purpose we generated proton air showers with an energy of 10$^{19}$~\eV{} 
 with the \texttt{CONEX}~\cite{Bergmann:2006yz} event generator. The resulting
longitudinal charged particle and energy deposit profiles were
subsequently fed into the atmosphere and detector simulation package~\cite{FDSim} of 
the Pierre Auger Observatory. 
The geometry and profile of the events in this simulated data sample was then reconstructed 
within the Auger offline software framework~\cite{offline}.\\
Only events satisfying basic quality selection criteria have been used in the analysis.
In order to assure a good reconstruction of the shower geometry, the angular length
of the shower image  on the camera was required to be larger than nine degrees. Moreover,
we only selected events with at least one coincident surface detector tank (so-called hybrid
geometry reconstruction~\cite{Sommers:1995dm}). 
Furthermore we rejected under-determined measured longitudinal profiles by demanding
an observed slant depth length of $\ge$ 300 \gcm{} and a reconstructed 
shower maximum within the field of view of the detector.\\
An example of a simulated event is shown in Fig.~\ref{fig:simulationExample}, illustrating
that the shape of the light curve at the detector can differ considerably from the
one of the energy deposit profile due to the scattered Cherenkov light
detected at late stages of the shower development. The reconstructed energy deposit curve, 
however, shows on average a good agreement with the generated profile.\\
Since longitudinal air shower profiles exhibit similar shapes when transformed from slant depth $X$
to shower age $s$ (see for instance~\cite{Giller:2005qz}),
a good test of the profile reconstruction performance is to compare
 the average generated and reconstructed energy deposit profiles as a function of $s$
normalized to the energy deposit at shower maximum.
As can be seen in Fig.~(\ref{fig:average}), the difference between these averages
is $\le$ 1.5\% and it can be concluded that the matrix method introduced here 
performs well in reconstructing air shower profiles without a prior assumption
about their functional shape. 
\section{Shower Age Dependence}
Due to the age dependence of the electron spectra $f_\mr{e}(E,s_i)$, the
Cherenkov yield  factors $Y^\mr{C}_i$ and the average electron energy deposits $\alpha_i$
depend on the depth of shower maximum, which is not known before the profile has been reconstructed. 
Fortunately, these dependencies are small: In the age range of importance for the
shower profile reconstruction ($s\in[0.8,1.2]$) $\alpha$ varies by only 
a few percent \cite{cher:nerling} and $Y^\mr{C}$ by less than 15\%~\cite{cher:giller}.
Therefore, a good estimate of $\alpha$ and $Y^\mr{C}$ can be obtained by setting
$s=1$ over the full profile or by estimating $\Xmax$ from the position maximum 
of the detected light profile.
 After the shower profile has been calculated with these estimates, 
 $\Xmax$ can be determined from the energy deposit profile 
and the profile can be re-calculated with an updated
Cherenkov-fluorescence matrix. The convergence of this procedure
is shown in Fig.~\ref{fig:convergence}. After only one iteration the $\Xmax$ (energy)
differs by less than 0.1 \gcm{} (0.1\%) from its asymptotic value.
Note that age dependent effects of the lateral spread 
of the shower on the image seen at the detector \cite{Gora:2005sq,Giller:2003ny},
though not discussed in detail here, have also been included in the 
simulation and reconstruction.
\section{Gaisser-Hillas Fit}
\begin{figure*}[t!]
  \begin{center}
       \subfigure[Energy.]{\label{fig:energyBias}
            \includegraphics[width=0.43\linewidth] {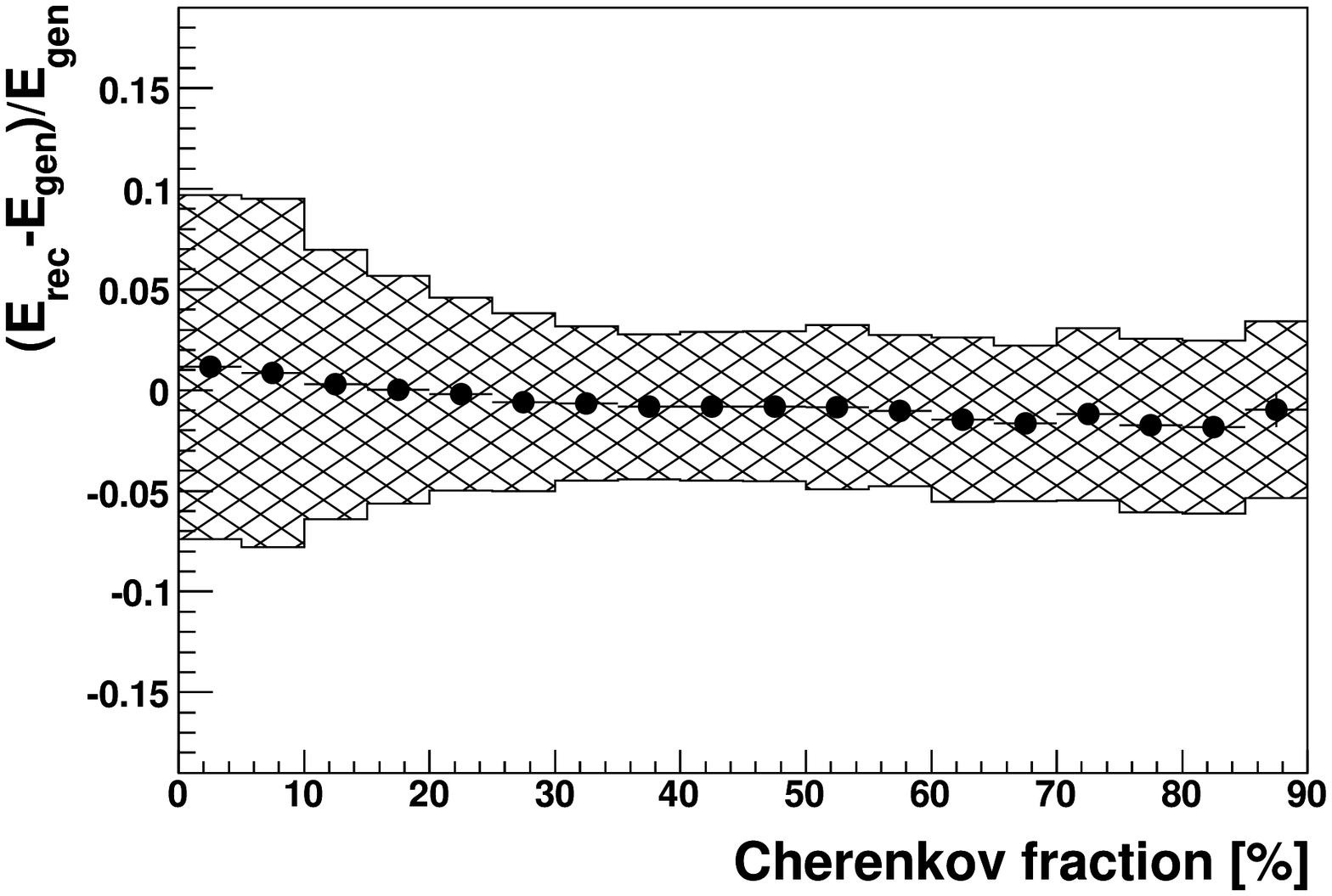}}
       \subfigure[$\Xmax$.]{\label{fig:xmaxBias}
            \includegraphics[width=0.43\linewidth] {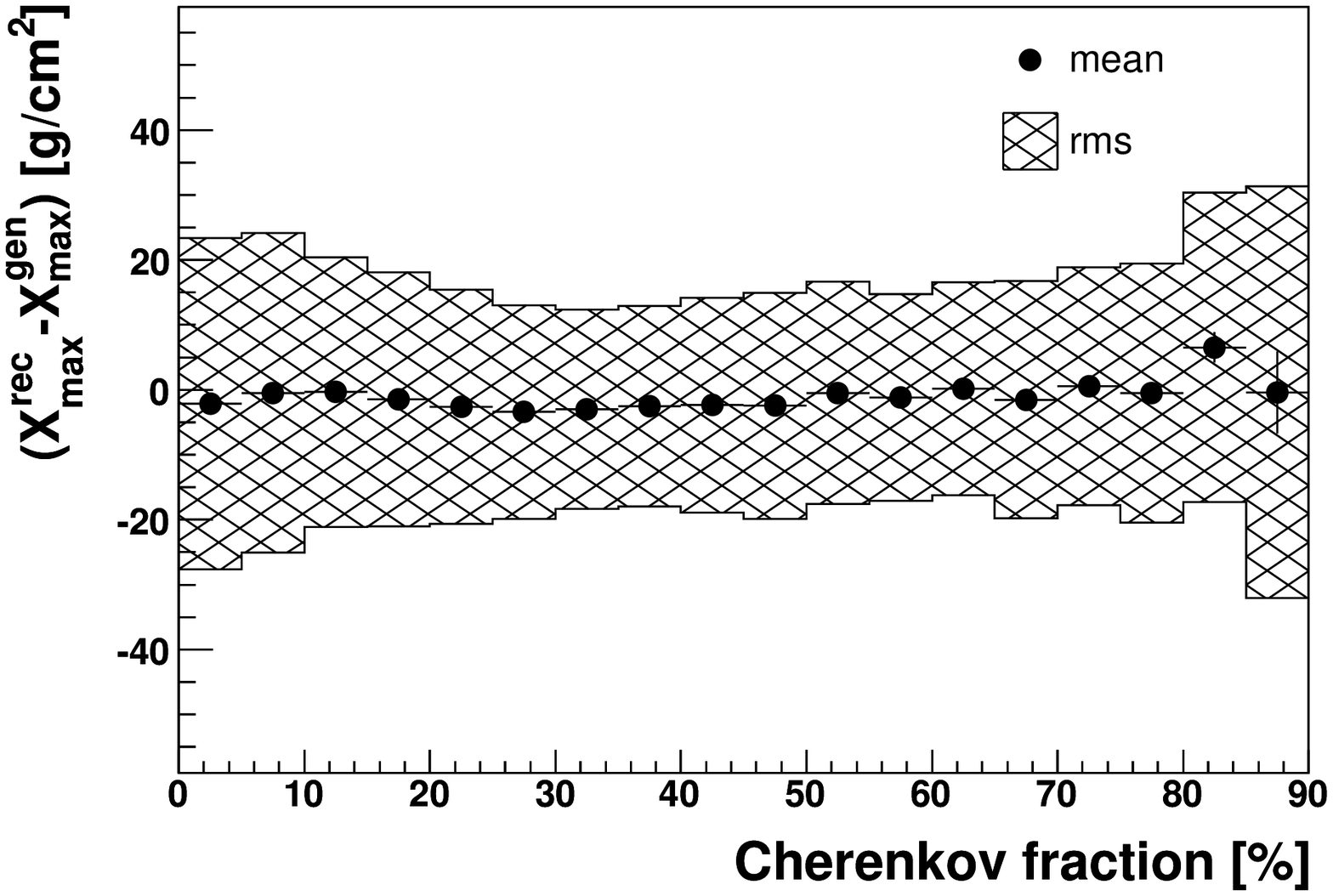}}
        \caption{Energy and $\Xmax$ reconstruction accuracy as function
                          of the amount of detected Cherenkov light.}
        \label{fig:bias}
   \end{center}
\end{figure*}
A knowledge of the complete profile is required for the calculation of the Cherenkov beam 
and the shower energy.
If due to the limited field of view of the detector only a part of the profile is observed, 
an appropriate function for the extrapolation to unobserved depths is needed. 
A possible choice is the Gaisser-Hillas 
function~\cite{ghfunc}
\be
  f_\mr{GH}(X)=\dEdX_\mr{max}  
      \left(\frac{X-X_0}{\Xmax-X_0}\right)^{(\Xmax-X_0)/\lambda} 
      e^{(\Xmax-X)/\lambda}\,,
  \label{eq:GH}
\ee
\begin{figure}[t]
  \begin{center}
    \includegraphics[width=\linewidth] {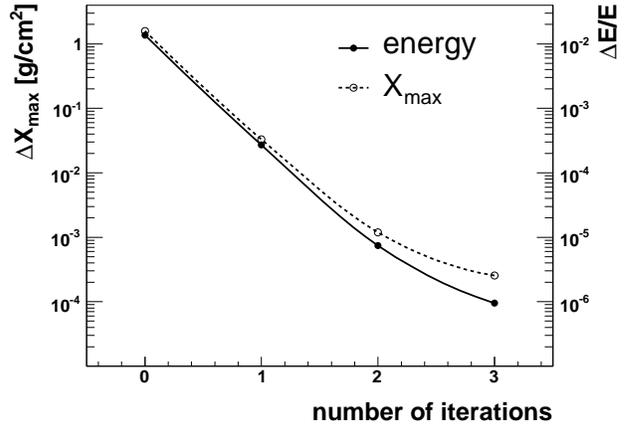}
        \caption{$\Xmax$ and energy difference with respect to the tenth shower age iteration.}
        \label{fig:convergence}
   \end{center}
\end{figure}
which was found to give a good description of measured longitudinal profiles~\cite{ghhires}. 
It has four free parameters: 
$\Xmax$, the depth where the shower reaches its maximum energy deposit $\dEdX_\mr{max}$
and two shape parameters $X_0$ and $\lambda$.\\
The best
set of Gaisser-Hillas parameters $\mbf{p}$ can be obtained by minimizing
the error weighted squared difference between the vector of function
values $\mbf{f_\mr{GH}}$ and $\widehat{\mbf{\dEdX}}$, which is
\be
 \chi^2_\mr{GH} = \left[\,\widehat{\mbf{\dEdX}}-\mbf{f(\mbf{p})}\right]^\mr{T}
           \,\mbf{V_\dEdX}^{\!\!\!-1}\,\left[\,\widehat{\mbf{\dEdX}}-\mbf{f(\mbf{p})}\right].
\ee
This minimization works well if a large fraction of the shower has been observed
below and above the shower maximum. If this is not the case, or even worse, if the shower
maximum is outside the field of view, the problem is under-determined, i.e.~the experimental
information is not sufficient to reconstruct all four Gaisser-Hillas parameters.
This complication can be overcome by constraining $X_0$ and $\lambda$ to their
average values $\langle X_0\rangle$ and $\langle\lambda\rangle$. The new 
minimization function is then the modified $\chi^2$
\be
  \chi^2=\chi^2_\mr{GH}+\frac{(X_0-\langle X_0\rangle)^2}{V_{X_0}}
          +\frac{(\lambda-\langle \lambda\rangle)^2}{V_\lambda} \,,
\label{eq:GHconstr}
\ee
where the variances of $X_0$ and $\lambda$ around their mean values are in the denominators.\\
In this way, even if $\chi^2_\mr{GH}$ is not sensitive to $X_0$ and $\lambda$, the minimization
will still converge. On the other hand, if the measurements have small statistical uncertainties and/or
cover a wide range in depth, the minimization function is flexible enough to allow for 
shape parameters differing from their mean values. These mean values can be determined 
from air shower simulations or, preferably, from high quality data profiles which can be
reconstructed without constraints.\\ Eq.~(\ref{eq:GHconstr}) can be
easily extended to incorporate correlations between $X_0$ and $\lambda$ and
the energy dependence of their mean values.
Air shower simulations indicate a small logarithmic energy dependence of the 
latter ($\le$ 25\% and 5\% per decade for $X_0$ and $\lambda$ respectively~\cite{lorenzo}).
In practice it is sufficient to use energy independent values determined at low energies,
because at high energies the number of measured points is large and thus the constraints 
do not contribute significantly to the overall $\chi^2$.

The accuracy of the reconstructed energy, obtained by integrating over the
Gaisser-Hillas function (see below), and that of the depth of shower maximum, are displayed in Fig.~\ref{fig:bias} as a function
of the relative amount of Cherenkov light.
Note that the good resolutions of $\approx$ 7\% and 20~\gcm{} are of course not
a feature of the reconstruction method alone, but depend strongly on the 
detector performance and quality selection. The mean values of difference
to the true shower parameters, however, which are
close to zero for both fluorescence and Cherenkov light dominated events, indicate
that both light sources are equally suited  to reconstruct the longitudinal development
of air showers.\\
A slight deterioration of the resolutions can be seen for events with a very small 
Cherenkov contribution of $<$ 10\%. 
Such showers are either inclined events which developed high in the atmosphere where
the light scattering probabilities are low or deep vertical showers, for which most
of the late part of the shower is below ground level. Both topologies result in
a somewhat worse resolution: The former correspond to larger than average distances to the detector 
and the latter to shorter observed profiles.

\section{Error Propagation}
A realistic estimate of the statistical uncertainties of important shower parameters
is desired for many purposes, like data quality selection cuts, or the comparison
between independent measurements like the surface and fluorescence detector measurements
of the Pierre Auger Observatory or the Telescope Array.
The uncertainties of $\dEdX_\mr{max}$, $\Xmax$, $X_0$ and $\lambda$ obtained after the minimization of 
Eq.~(\ref{eq:GHconstr}), reflect only
the statistical uncertainty of the
light flux, which is why these errors will be referred to as 'flux uncertainties' 
($\sigma_\mr{flux}$) in the following. Additional uncertainties arise from the uncertainties
on the shower geometry ($\sigma_\mr{geo}$), atmosphere ($\sigma_\mr{atm}$) 
and the correction for invisible energy ($\sigma_\mr{inv}$).

\subsection{Flux uncertainty of the calorimetric energy}
Even with the flux uncertainties of the Gaisser-Hillas parameters
it is not straightforward to calculate the flux uncertainty of the
calorimetric energy, which is given by the integral over
the energy deposit profile:
\be
  E_\mr{cal} = \int_0^\infty f_\mr{GH}(X)\, \mr{d}X\;.
  \label{eq:Eem}
\ee
To solve this integral one can substitute
\be
   t=\frac{X-X_0}{\lambda} 
\;\;\;\;\;\;\text{and}\;\;\;\;\;\;
   \xi=\frac{\Xmax-X_0}{\lambda}
\ee
in the Gaisser-Hillas function Eq.~(\ref{eq:GH}) to get
\be
  f_\mr{GH}(t)=\dEdX_\mr{max}  \left(\frac{e}{\xi}\right)^\xi\, e^{-t}\, t^\xi,
\ee
which can be identified with a Gamma distribution. Therefore, the above integral is given 
by
\be
   E_\mr{cal} = \lambda\,\dEdX_\mr{max} \left(\frac{e}{\xi}\right)^\xi\, \Gamma(\xi+1)\,, 
\ee
where $\Gamma$ denotes the Gamma-function. Thus, instead of doing a tedious error propagation
to determine the statistical uncertainty of $E_\mr{cal}$ one can simply use it directly 
as a free parameter in the fit instead of the conventional factor $\dEdX_\mr{max}$:
\be
  f_\mr{GH}(t)=\frac{E_\mr{cal}}{\lambda\,\Gamma(\xi+1)}\, e^{-t}\, t^\xi
  \label{eq:ghwithe}
\ee 
In this way, $\sigma_\mr{flux}(E_\mr{cal})$ is obtained directly from the $\chi^2+1$ contour of 
Eq.~(\ref{eq:GHconstr}).
\subsection{Geometric uncertainties}
\label{sec:geomerr}
Due to the uncertainties on the shower geometry, the distances $r_i$ to
each shower point are only known within a limited precision and correspondingly the energy deposit
profile points are uncertain due to the transmission factors $T(r_i)$
and geometry factors $1/(4\,\pi\,r_i^2)$. Furthermore, the uncertainty of the
shower direction, especially the zenith angle $\theta$, affects
the slant depth calculation via $X_\mr{slant}=X_\mr{vert}/\cos\theta$ and thus $\Xmax$. 
Finally, the amount of direct and scattered Cherenkov light depends on the shower geometry, too,
via the angles $\beta_i$.\\

The algorithms used to reconstruct the shower geometry from fluorescence detector data  
usually determine the
following five parameters~\cite{Baltrusaitis:1985mx}, irrespective of whether the detectors
operate in monocular, stereo or hybrid mode:
\be
   {\bf\alpha}=\{\theta_\mathrm{SDP},\Phi_\mathrm{SDP},T_0,R_p,\chi_0\}.
\ee
$\theta_{SDP}$ and $\Phi_{SDP}$ are the angles of the normal vector of a plane spanned by the
shower axis and the detector (the so called shower-detector-plane), $\chi_0$ denotes the angle
of the shower within this plane and $T_0$ and $R_p$ are the time and distance of the shower
at its point of closest approach to the detector.\\
For any function $q({\bf\alpha})$ standard error propagation yields the geometric uncertainty
\be
    \sigma_\mr{geom}^2(q)=\sum_{i=1}^5\sum_{j=1}^5 \frac{\mr{d} q}{\mr{d} \alpha_i}\frac{\mr{d} q}{\mr{d} \alpha_j} V^\alpha_{ij},
     \label{eq:geomerrprop}
\ee
where $\bf V^\alpha $
denotes the covariance matrix of the axis parameters.
As the calorimetric energy and $\Xmax$ depend non-trivially on the shower geometry, the above 
derivatives need to be calculated numerically, i.e. by repeating the profile 
reconstruction and Gaisser-Hillas fitting for the ten new geometries given by $\alpha_i\pm\sqrt{V^\alpha_{ii}}
\equiv \alpha_i\pm\sigma_i$ 
to obtain
\bea
    \Delta_i &=&\frac{\mr{d} q}{\mr{d} \alpha_i}\sigma_i \nonumber \\
             &\approx& \frac{1}{2}[\, q(\alpha_i + \sigma_i)  - q(\alpha_i - \sigma_i ) \,],
    \label{eq:errpropdelta}
\eea
with which Eq.~(\ref{eq:geomerrprop}) reads as
\be
    \sigma_\mr{geom}^2(q)=\sum_{i=1}^5\sum_{j=1}^5 \Delta_i \Delta_j \rho_{ij}
     \label{eq:simpleerrprop}
\ee
where
\be
    \rho_{ij}=\frac{V^\alpha_{ij}}{\sqrt{V^\alpha_{ii}V^\alpha_{jj}}}
\ee
denote the correlation coefficients of the geometry parameters $\alpha_i$ and $\alpha_j$.
\begin{figure}[t]
  \begin{center}
       \subfigure[Energy.]{\label{fig:energyPull}
            \includegraphics[width=0.48\linewidth] {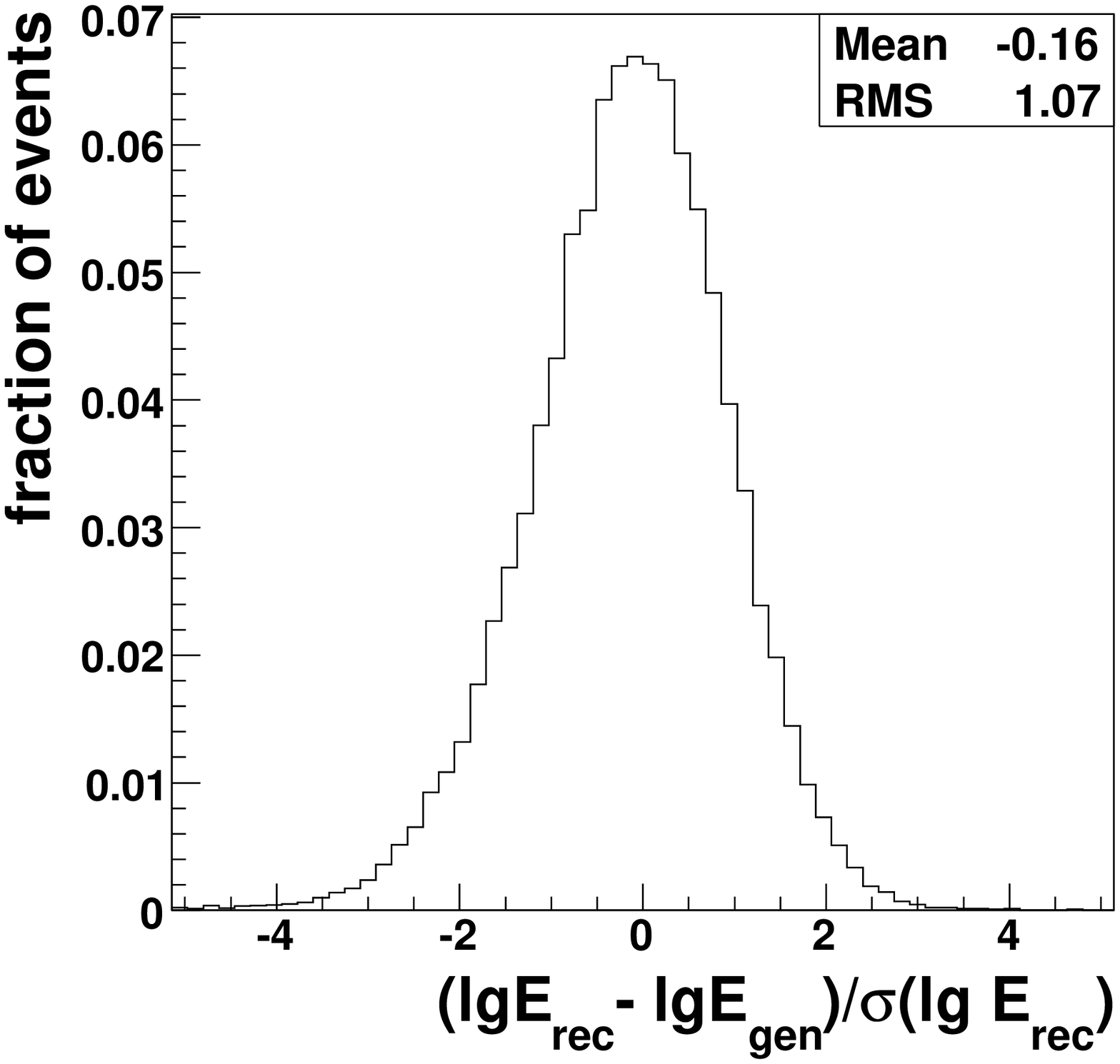}}
       \subfigure[$\Xmax$.]{\label{fig:xmaxPull}
            \includegraphics[width=0.48\linewidth] {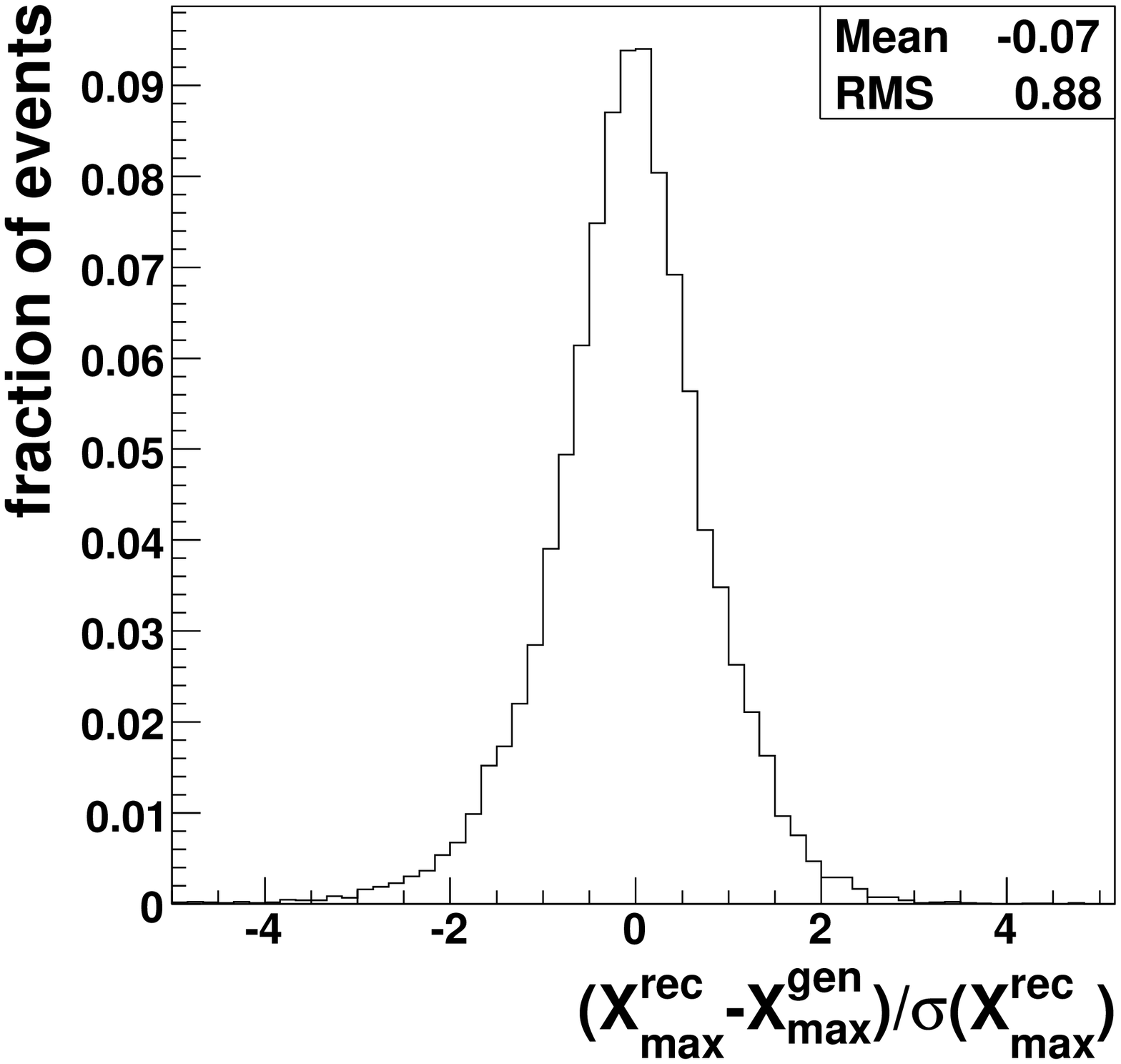}}
        \caption{Pull distributions}
        \label{fig:pull}
   \end{center}
\end{figure}
\subsection{Atmospheric uncertainties}
Whereas the Rayleigh attenuation is a theoretically well understood process,
the molecular density profiles and aerosol content of the atmosphere vary
due to environmental influences and need
to be well monitored in order to determine the slant depth and transmission
coefficients needed for the profile reconstruction. Uncertainties in
these measured atmospheric properties (see for instance \cite{BenZvi:2007it,Keilhauer:2005ja})
can be propagated in the same way as the geometric uncertainties
by determining the one sigma shower parameter deviations via Eq.~(\ref{eq:errpropdelta}).
\subsection{Invisible energy}
Not all of the energy of a primary cosmic ray particle ends up
in the electromagnetic part of an air shower. Neutrinos escape
undetected and muons need long path lengths to fully release their energy. This 
is usually accounted for by multiplying the calorimetric energy, Eq.~(\ref{eq:Eem}), 
with a correction factor $f_\mr{inv}$ determined from shower simulations to
obtain the total primary energy
\be
   E_\mr{tot}=f_\mr{inv}\,E_\mr{cal}.
   \label{eq:etot}
\ee
The meson decay probabilities, and thus the amount of neutrino and muon
production, decrease with energy, therefore  
$f_\mr{inv}$ depends on the primary energy. For instance, in
\cite{invE:barbosa} it is parameterized as 
\be
    f_\mr{inv}=\left(a+b E_\mr{cal}^c\right)^{-1},
\ee
where $a$, $b$ and $c$ denote constants depending on the primary composition and
interaction model assumed\footnote{Note that here only the  {\itshape statistical} 
uncertainties of the invisible energy correction are discussed. For an
estimate on the related {\itshape systematic} uncertainties see~\cite{icrc05:tanguy}}.
 This energy dependence needs to be
taken into account when propagating the calorimetric
energy uncertainty to the
total energy uncertainty.

Due to the stochastic nature of air showers, the correction factor is 
subject to shower-to-shower fluctuations. The statistical uncertainty of $f_\mr{inv}$
was determined in~\cite{icrc05:tanguy}
and can be parameterized as follows:
\be
   \sigma(f_\mr{inv})\approx 1.663\cdot 10^6 \cdot \lg(E_\mr{tot}/\eV)^{-6.36}.
\ee
Typical values are 2.5\% at 10$^{17}$~\eV{} and 0.9\% at 10$^{20}$ \eV{}.\\

\subsection{Total statistical uncertainty}
Summarizing the above considerations, the statistical variance of the total energy
is 
\bea
   \sigma_\mr{stat}^2(E_\mr{tot})&=&E_\mr{tot}^2\,\sigma^2(f_\mr{inv}) \nonumber \\
                      &&+\left(\frac{\mr{d}f_\mr{inv}}{\mr{d}E_\mr{cal}}E_\mr{cal}
                       +f_\mr{inv}\right)^2
                       \sum_i \sigma_i^2(E_\mr{cal}),
    \label{eq:eerr}
\eea
where $i$ runs over the geometric, atmospheric and flux uncertainties.
Since the invisible energy correction does not affect the depth of shower maximum, its
uncertainty is simply given by
\be
    \sigma_\mr{stat}(\Xmax)=\sqrt{ \sum_i \sigma_i^2(\Xmax)}\,.
    \label{eq:xmaxerr}
\ee

Again we use simulated events to verify the validity of the above considerations. 
The pull distributions of the reconstructed energy and
shower maximum, shown in Fig.~\ref{eq:xmaxerr}, both have a width of approximately one, 
which means that the total uncertainties from Eqs.~(\ref{eq:eerr}) and~(\ref{eq:xmaxerr})
are good estimators for the actual event-by-event measurement uncertainties.

\section{Conclusions and Outlook}
In this paper a new method for the reconstruction of longitudinal air shower
profiles was presented. With the help of simulations we have shown that
the least square solution yields robust and unbiased results and that
uncertainties of shower parameters can be reliably calculated for
each event.\\
Events with a large
Cherenkov light contribution are currently usually rejected during
the data analysis (see for instance \cite{Abbasi:2004nz,AbuZayyad:2000ay})
However, as we have shown, there is no justification for rejecting such showers, once experimental systematic uncertainties are well understood.
Because events with a large Cherenkov contribution have different systematic uncertainties to those dominated by fluorescence light, both event
classes can be compared to study their compatibility.\\
At energies below 10$^{17.5}$ \eV{}, where new projects \cite{heat,tale} 
are planned to study the transition from galactic to 
extragalactic cosmic rays, events with a large fraction of direct Cherenkov
light will dominate the data samples, because the amount of
light, and thus trigger probability, of these events is much larger than
that of a fluorescence dominated shower. If at these energies it is
still possible to measure an accurate shower geometry, the
fluorescence detectors should in fact be used as
Cherenkov-Fluorescence telescopes.\\

\section*{Acknowledgments}
The authors would like to thank their colleagues from the Pierre Auger
Collaboration, in particular Frank Nerling and Tanguy Pierog, for
fruitful discussions.  

\newpage

\bibliography{profile}
\bibliographystyle{elsart-num}
\end{document}